\documentclass[twocolumn,prd,showpacs,nofootinbib,a4]{revtex4}
\usepackage{graphicx}

\def\d{{\rm d}}
\begin{document}

\title{Astrophysical limitations to the identification of dark matter:\\ indirect neutrino signals vis-{\`a}-vis direct detection recoil rates}
\author{Pasquale D.~Serpico}
\affiliation{Physics Department, Theory Group, CERN, CH--1211 Geneva 23, Switzerland}
\affiliation{LAPTH, UMR 5108, 9 chemin de Bellevue - BP 110, 74941 Annecy-Le-Vieux, France}
\date{\today}
\author{Gianfranco Bertone}
\affiliation{Institut d'Astrophysique de Paris, France. UMR7095-CNRS UPMC,
98bis Boulevard Arago, 75014 Paris, France}
\affiliation{Institute for Theoretical Physics, University of Zurich, 8057 Zurich, Switzerland}

\begin{abstract}
A convincing identification of dark matter (DM) particles can probably be achieved only through a combined analysis of different detections strategies, which provides an effective way of removing degeneracies in the parameter space of DM models. In practice, however, this program is made complicated by the fact that different strategies depend on different physical quantities, or on the same quantities but in a different way, making the treatment of systematic errors rather tricky. We discuss here the uncertainties on the recoil rate in direct detection experiments and on the muon rate induced by neutrinos from dark matter annihilations in the Sun, and we show that, contrarily to the local DM density or overall cross section scale, irreducible astrophysical uncertainties affect the two rates in a different fashion, therefore limiting our ability to reconstruct the parameters of the dark matter particle. By varying within their respective errors astrophysical parameters such as the escape velocity and the velocity dispersion of dark matter particles, we show that the uncertainty on the relative strength of the neutrino and direct-detection signal is as large as a factor of two for typical values of the parameters, but can be even larger  in some circumstances.
\end{abstract}
\pacs{ 95.35.+d, 14.60.Lm \hfill CERN-PH-TH/2010-139, LAPTH-020/2010}
%Dark matter, standard model neutrinos

\maketitle
\section{Introduction}

The detection of Dark Matter (DM) particles in accelerators or with direct and indirect searches would be of paramount importance for particle physics and cosmology. A convincing identification of DM, however, will require a combined analysis of different experimental strategies (see Refs. ~\cite{Jungman:1995df,Bergstrom:2000pn,Bertone:2004pz,book} for recent reviews). From LHC data alone, for instance, it will be difficult to reconstruct the relic density of the newly discovered particles (if any), even if one assumes to know the underlying theory (e.g. Ref. \cite{Baltz:2006fm}. See also \cite{Nath:2010zj} and references therein). However, combining LHC data e.g. with direct detection data significantly improves the 
reconstruction~\cite{Bertone:2010rv}. One could cite several other examples of complementarity of different detection techniques, such as the combination of different targets in direct detection experiments~\cite{Bertone:2007xj,Drees:2008bv,Peter:2009ak,Shan:2010qv}, or multi-wavelength analyses in indirect detection experiments~\cite{Profumo:2010ya}. 

Although the combination of different detection strategies provides an effective way of removing degeneracies in the parameter space of DM model, a combined analysis is made complicated by the fact that one needs to put together results that depend on different physical quantities, or on the same quantities but in a different way, a circumstance that makes the treatment of systematic errors (i.e. depending on the underlying model assumptions) rather problematic. 

The statistical and systematic uncertainties affecting the reconstruction of DM particle from various experiments have been discussed widely in literature. One recent example is the the uncertainty on the local DM density, affecting both the rate of nuclear recoils in direct detection experiments, and the flux of upgoing muons induce by DM annihilations in the Sun. The statistical error on this quantity, based on dynamical constraints of various tracers in the Galaxy, have been discussed in Refs. \cite{dyn1,dyn2,dyn3,dyn4} and more recently in \cite{Catena:2009mf,Strigari:2009zb} (see also Ref. \cite{Salucci:2010qr}). The systematic error due to the poor knowledge of the DM density profile in the Galaxy, and in particular to its possibly triaxiality has been discussed in Ref.~\cite{Zurich10}. Another recent example is the solar model dependence of the DM-induced neutrino signal~\cite{Ellis:2009ka}.

Here we want to assess a minimal level of uncertainty in the normalization of the neutrino flux from the Sun (and, a fortiori, the Earth) due to the ignorance on crucial astrophysical parameters, in particular concerning the DM phase space distribution.
While the impact of these uncertainties for direct dark matter searches has been long acknowledged (see e.g.~\cite{Belli:2002yt}) and it is still actively investigated (see for example~\cite{Kuhlen:2009vh,Vogelsberger:2008qb}) the impact for indirect neutrino detection has been investigated less extensively. Occasionally, it has been noted that some effects (e.g. clumpiness and dark disk, see Sec.~III.5 below) may enhance the ``baseline'' flux, but little attention has been paid to the robustness of the baseline flux from an ordinary, smooth halo.
Also, it is of interest to address the question of what would be the uncertainty on the prediction for the indirect neutrino signal, should a direct detection be used for normalization. Finally, since this  uncertainty turns out to be the dominant one, its value also sets the accuracy needed for several nuclear and particle physics input  parameters, derived either empirically or via theoretical computations.

Here, we study this issue also in light of recent results from numerical simulations. The structure of the article is the following: in Sec. \ref{theory} we review the dependence of the recoil and muon rates on key astrophysical parameters. In Sec.~\ref{params} we discuss the systematic uncertainties inevitably associated with these parameters, including time-dependent effects. 
Sections~\ref{theory}  and \ref{params} are intended mostly as reviews of the essential concepts
and quantities and to introduce our notation.
In Sec. \ref{results} we discuss the impact of uncertainties on four variables discussed in~Sec.~\ref{params} on theoretical predictions for the recoil and neutrino rates. We also illustrate the qualitative changes in the absolute and relative uncertainties as a function of the dark matter mass as well as direct detection experiment
target material and energy-threshold. Finally, in Sec.\ref{conclusions} we discuss the results and present our conclusions.

\section{Theoretical Setup}
\label{theory}

The theory of DM experiments has been widely discussed in literature. We summarize in this section the dependence of the predicted detection rates on some key parameters. 

\subsection{Recoil rate in direct detection experiments}
We start from direct detection experiments, for which, unless otherwise stated, we follow the notation of~\cite{Belanger:2008sj}. Apart for numerical constants of order one, the differential number of recoil events 
per unit time and target mass writes 
\begin{equation}
\frac{\d R}{\d E}\sim \frac{\sigma_{0,A}}{\mu_{\chi,A}^2}\frac{\rho_\odot}{m_{\chi}}F_A^2(E) \int \d v \frac{ f_1(v)}{v}\Theta(v-v_{\rm min})\,\label{dRdE}
\end{equation} 
where $\mu_{\chi,A}=m_\chi\,M_A/(m_\chi+M_A)$, $v_{\rm min}=\sqrt{\frac{E\,M_A}{2\,\mu_\chi^2}}$
and
\begin{equation}
f_1(v)\equiv \int \d \Omega\, v^2\, f({\bf v})\Rightarrow \int_0^\infty \d v\, f_1(v)=1\,,
\end{equation} 
where $m_\chi$ is the DM mass, $M_A$ the target atom mass, $\sigma_{0,A}$ the DM-nucleus cross-section, $F_A$ the form factor, $\rho_\odot$ the local density of DM, and $f({\bf v})$ is the 3D velocity distribution. Note that $f_1(v)$ denotes the velocity distribution of DM particles {\it at the Earth}. If one has the distribution of velocity in the ``halo rest frame'', in general one has to correct for: i) time-dependent velocity of the detector with respect to this halo, including Earth motion with respect to the Sun and Solar motion with respect to the halo; ii) Gravitational effects, including in general both the gravitational focusing of unbound particles
(see e.g.~\cite{Alenazi:2006wu} and refs therein) and the population of gravitationally bound
DM particles (see~\cite{Lundberg:2004dn,Peter:2009mi} and refs. therein).

We shall not enter the details of the form factor dependence to avoid introducing too
many independent parameters: for our purposes, we shall use the parameterization in terms
of the nuclear mass from~\cite{Jungman:1995df} 
\begin{equation}
F_A^2(E)=\exp(-m_A\,R_A^2\,E/3)\,
\end{equation}
with $R_A=1\, {\rm fm}[0.3+0.91\,\sqrt[3]{m_{A,{\rm GeV}}}]$. The overall rate $R$ of interest here is
obtained by integrating  Eq.~(\ref{dRdE}) over energy, starting from the threshold $E_T$, which is the minimum recoil-energy detectable, which is experiment-dependent but usually of the order of $\sim$ 10 keV. 
So, the input parameters to specify in this case are: $\{m_A,E_T, m_\chi,\sigma_{0,A},{\bf y}\}$,
where ${\bf y}$ denotes collectively the astrophysical parameters. 

\subsection{Muon rate induced by neutrinos from DM annihilations in the Sun}

As mentioned above, the detection of a high-energy neutrino flux from the center of the Sun or the Earth would provide a  very convincing evidence for DM annihilations. The rate of events in neutrino telescopes depends on the number of particles that are captured, and on their fate inside the star or planet. 
  
The number of DM particles captured in the Sun/Earth obeys the following time evolution equation
\begin{equation}
\dot{N}=C-C_A N^2
\end{equation}
where $C$ is the capture rate and $C_A$ (see below) regulates DM annihilations. 
{\it If} both coefficients are constant, solving for $N(t)$ one can derive 
\begin{equation}
\Gamma_A(t)=\frac{C}{2}\tanh^2\left(\frac{t}{\tau_{\rm eq}}\right)\,,\:\:\:\tau_{\rm eq}=(C\,C_A)^{-1/2}\,.\label{eqtime}
\end{equation}
In the limit where steady state is reached within timescales much shorter than the lifetime of the Sun, $t_\odot\simeq 4.6\times 10^9\,$yr, one has $\dot{N}=0$, $N=\sqrt{C/C_A}$,
and the annihilation rate writes
\begin{equation}
\Gamma_A=\frac{C_A}{2} {N_{\rm eq}}^2=\frac{C}{2}\,. 
\end{equation}
In this regime, the normalization of the signal only depends on $C$. 
More generally, the present value $\Gamma_A(t_\odot)$ depends also on $\tau_{\rm eq}$, i.e. on $C_A$. $C_A$ can be written in terms of effective volumes $V_{1,2}$, as
\begin{equation}
C_A=\langle \sigma_A v\rangle\frac{V_2}{V_1^2} \,,\:\:\:V_j\simeq\left(\frac{3\,m_{\rm Pl}^2T_\circ}{2\,j\, m_\chi \rho_{\circ}}\right)^{3/2}
\end{equation}
with $T_\circ\,,\rho_\circ$ respectively the central temperature and density of the body under consideration~\cite{Jungman:1995df}. Note that the above formula already has an uncertainty due to the only approximate assumption of thermalization, homogeneous conditions of the core, etc., some of which are known to fail in some circumstances, see e.g.~\cite{Peter:2008sy}. In the other useful limit $\tau_{\rm eq}\gg t_\odot$, 
\begin{equation}
\Gamma_A(t_\odot)\approx 0.5\,C^2 C_A\,t_\odot^2\,.\label{gammalongeq}
\end{equation}
This has the following implications for the large $\tau_{\rm eq}$ regime: i) the signal is quadratic in $C$; ii) there is an additional dependence on $C_A$ (i.e.  linear in $\langle \sigma_A v\rangle(\frac{m_\chi \rho_{\circ}}{T_\circ})^{3/2}$) as well as a quadratic one on $t_\odot$. Even neglecting particle physics unknowns, this means that additional astrophysical uncertainties enter.  
For the Sun, different models (see e.g. Ref.~\cite{serenelli,Dogan:2010my}) yield different predictions for these parameters at the $\sim 1\%$ level.
For the Earth, uncertainties are much larger. For example, according to~\cite{earth}, 
EarthÕs temperature at the {\it inner core boundary} is estimated at $5650\pm 600\,$K.We note incidentally that despite these uncertainties, it is possible to obtain useful constraints on DM 
particles with very large cross-sections, higher than those excluded by direct detection experiments, by requiring that they do not over-heat the Earth~\cite{Mack:2007xj}.

Focusing on $C$, we shall first remind that for most particle physics models (including Kaluza-Klein and most neutralino DM models), the capture in the Sun is actually dominated by spin-dependent interactions on hydrogen~\cite{Bertone:2004pz}. This implies among others that we can consider the form factor~$\approx 1$, which greatly simplifies the formulae.
Although exceptions do exist, in the following we are interested in discussing the uncertainties induced
by the astrophysical (rather than nuclear/particle physics) input, so we shall limit ourselves to this single contribution in the capture.

From rewriting Eqs.~(2.8,2.13) of~\cite{Gould:1987ir} in our notation we get
\begin{eqnarray}
C&=&\sigma_{0,p}\frac{\rho_\odot\,\epsilon_p\,M_\odot}{m_\chi\,m_p}\times \nonumber\\
&&\int_0^1 \d {\cal M}\,\nu^2({\cal M})\int_0^{u_{\rm max}} \d u \frac{f_1(u)}{u}\left[1-\frac{u^2}{u_{\rm max}^2}\right]\,,
\end{eqnarray}
where
\begin{equation}
u_{\rm max}({\cal M})\equiv \frac{\sqrt{4\,m_\chi\,m_p}}{m_\chi-m_p}\nu({\cal M}) \label{umax}
\end{equation}
and $\nu$ is the escape velocity from the unit shell volume considered, in turn depending on the distance from the center $r$; to a good approximation~\cite{Gould:1992,Ellis:2009ka},
\begin{eqnarray}
\nu^2(r) = v_{\circ}^2-{\cal M}(r)(v_{\circ}^2-v_s^2)\,,\nonumber\\
v_{\circ}\simeq 1355\,{\rm km/s}\,,\:v_s\simeq 818\,{\rm km/s}\,
\end{eqnarray} 
where ${\cal M}(r)$ is the mass within the radius $r$, in units of the solar mass. To derive the above equation for the capture rate, we replaced the Sun with a pure hydrogen sphere and introduced the factor $\epsilon_p<1$ to account for the fact that only a fraction of the Sun is made of hydrogen. What is crucial here
is the approximation that the density of hydrogen traces the total one of the Sun, which amounts to
a small ($\sim 8\%$) overestimate of the capture~\cite{Ellis:2009ka}. Note that the consequences of our approximations are further softened by the fact that we are only concerned with {\it relative} response of the observables to variations in the astrophysical inputs, so that overall scale mismatches are of no importance here.

The main input parameters in this case are thus $\{m_\chi,\sigma_{0,p},{\bf y}\}$.
Further simplifications might be obtained by specifying the function $f_1$, but we do not proceed further along this path, since we want to keep some generality concerning $f_1$, whose key parameters are described in the following section. 
 
\begin{figure}[t]
\begin{center}
\begin{tabular}{c}
\includegraphics[angle=0,width=0.49\textwidth]{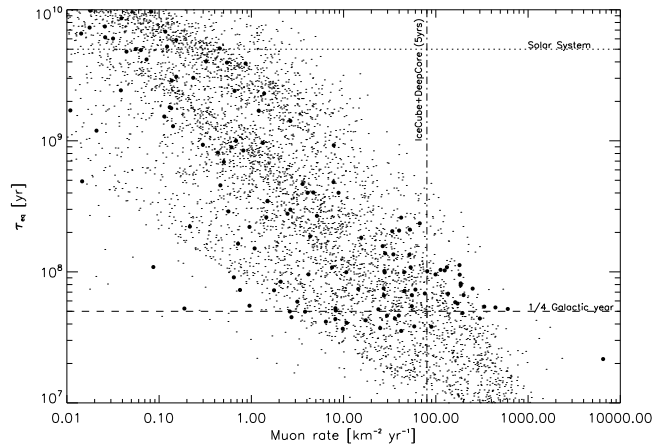}
\end{tabular}
\end{center}
\caption{Timescale to reach equilibrium between capture and annihilation of DM particles in the Sun (Eq.~(\ref{eqtime})) as a function of the muon rate on a terrestrial detector for DM models in a pMSSM scenario, as implemented in DarkSUSY \cite{Gondolo:2004sc}. Solid circles correspond to models that fulfil the constraints on the cosmological relic abundance.The two horizontal lines
correspond to the solar system  age and 1/4 of the solar revolution time in the Galaxy. The vertical line shows the approxmate sensitivity of IceCube plus DeepCore in 5 years of data taking  \cite{Halzen:2009vu}.
\label{fig:eq}}
\end{figure}

Neutrino experiments such as IceCube are mostly sensitive to the flux of up-going muons induced by high-energy neutrinos. 
The rate of $\mu$'s induced by a neutrino flux $\Phi_{\nu}(E_{\nu})$
is given by
\begin{equation}
\Gamma_\mu = \int_{E_{\mu}^{\rm thr}}^{m_\chi} \d E_{\nu}
  \int_{0}^{y_{\nu}} \d y\,
   A(E_{\mu})P_{\mu}(E_{\nu},y) \Phi_{\nu}(E_{\nu})
\label{rate}
\end{equation}
where $y_{\nu}=1-E_{\mu}^{\rm thr}/E_{\nu}$ and $E_{\mu}^{\rm{thr}}$ is the muon threshold energy of the experiment. $A(E_{\mu})$ is the effective area of the detector and $P_{\mu}(E_{\nu},y)$ is the probability that a neutrino of energy 
$E_{\nu}$ interacts with a nucleon producing a muon of energy $E_{\mu} \equiv (1-y)E_{\nu}$ 
above the detector threshold energy, and can be estimated as 
 \begin{equation}
P_{\mu}(E_{\nu},y) \simeq\frac{\tilde{\rho}}{m_p}\,\tilde{R}(E_{\mu},\, E_{\mu}^{\rm thr}) \,
\sigma(E_{\nu},y)
\end{equation}
where $\tilde{\rho}$ is the density of the medium (typically water/ice),  $\tilde{R}(E_{\mu},\, E_{\mu}^{\rm thr})$ is the muon range in that medium, 
{\it i.e.} the distance 
traveled by muons before their energy drops below 
$E_{\mu}^{\rm thr}$,  and 
$\sigma(E_{\nu},y)\equiv \d\sigma_{CC}^{\nu N}(E_{\nu},y)/\d y$ is the differential
cross section for neutrino--nucleon charged--current scattering
(for further details see e.g. Ref.~\cite{Bertone:2004pz} and 
references therein).
In our case, the neutrino flux is simply given by 
\begin{equation}
\Phi_\nu = \frac{\Gamma_A}{4 \pi D^2} \frac{dN_\nu}{dE}
\end{equation}
where $dN_\nu/dE$ is the neutrino spectrum per annihilation and $D$ is the distance of the Sun to the Earth (Astronomical Unit).

We show in Fig.~\ref{fig:eq} the timescale to reach equilibrium between capture and annihilation as a function of the muon rate on a terrestrial detector for DM models in a pMSSM (``Phenomenological Minimal Super-Symmetric Standard Model'') scenario, as implemented in DarkSUSY~\cite{Gondolo:2004sc}. For reference, we also show with horizontal curves two characteristic timescales of the solar system: its age, and 1/4 of the revolution time around the Galactic center $\tau_{\rm gal}$, i.e. the time to span a quadrant. The vertical line shows the approxmate sensitivity of IceCube plus DeepCore in 5 years of data taking (e.g. Ref. \cite{Halzen:2009vu}).  It is clear that most interesting model in this scenario have  $\tau_{\rm eq}\in (0.2; 1)\,\tau_{\rm gal}$, which makes ``galactic'' effects potentially important. Also note that for the Earth equilibrium is rarely reached in interesting models, see for example Fig. 18 in~\cite{Lundberg:2004dn}.

%%%
\section{Astrophysical uncertainties}
\label{params}
%%%
In general, both DM velocity and density distributions crucially affect the  the predictions for observables.  Since in the literature most attention has been paid to the {\it common} uncertainty coming from the local normalization value of the DM density, here we shall be mostly concerned with astrophysical quantities affecting {\it differently} direct and indirect signals. In particular, in the following we shall treat mostly velocity distribution uncertainties (subsections III.1,2,3). In subsections III.4,5, however,  we shall also comment on the role of time-dependent effects, which possibly introduce a "relative bias" between local density in direct experiments and long-term, averaged properties probed via capture. This is a subject rarely mentioned in the literature. 

Preliminarly to a more detailed discussion, it is worth recalling some facts about a crucial ingredient to make predictions for, or intepret, DM experiments:  $v_c(R_\odot)$, the velocity that a test-mass would have on a circular orbit in the Galactic Plane at the solar distance $R_\odot$ from the GC~\footnote{Actually, $v_c(R_\odot)$ only matters with the Ansatz that the local phase-space
density is dominated by a smooth halo component that is at rest in a Galactocentric frame. Also implicit throughout this paper is that only isotropic velocity distributions are considered. Evidently, more realistic assumptions can only inflate the error budgets with respect to those considered here.}.  
Note that, to a good approximation, the Milky Way mass distributions is axisymmetric: one can find a  cylindrical coordinate system $\{R,\phi,z\}$ centered on the GC where the potential is independent of $\phi$, with the $z=0$ plane dubbed ``Galactic Plane'', (see below the discussion of deviations from axial symmetry). For a particle having  initial position at a distance $R$ from the GC, with no radial and no vertical component of the velocity vector, the motion reduced to one in an attractive central potential. For any distribution of mass one can find a value of the azimuthal velocity  $v_c(R)$ which keeps the particle on a circular orbit at distance $R$, known as {\it circular velocity}. Formally,  in analogy with the formula for spherically symmetric systems, one can always write 
\begin{equation}
\frac{G\,M}{R^2}=\frac{v_c(R)^2}{R}\,,
\end{equation} 
where $M$ is a parameter depending  on $R$ as well as the form of the potential. Actually, for axisymmetric potentials, the mass parameter $M$ represents the mass enclosed within the radius $R$ only for Mestel's disks, i.e. for disks whose surface density scales like $\Sigma(R) \propto 1/R$, while in general $v_c$ will depend also on the mass {\it outside} $R$. For instance, in the case of an exponential disk, the above equation underestimates $v_c$ by up to 15\% for radii larger than the scale radius of the disk~\cite{BT}.  
Anyway, real stars have orbits varying in all coordinates. For stars having initial velocities in the plane $z=0$, one has exactly $\langle v_z\rangle=0$, while the orbit has a ``rosetta'' shape in the $\{R,\phi\}$ plane, with $R$ varying between a minimum and a maximum value. If the star moves off-plane, then another oscillation takes place above/below the plane (for more details, see e.g. Chap. 3 in~\cite{BT}).
So, in general to derive $v_c$ requires identifying the circular motion of the ``local standard of rest'' (LSR), with respect to which actual stars in our neighborood (including the Sun) possess relative motions.
In~\cite{Catena:2009mf}, this was evaluated from a compilation of data to be $v_c=245\pm 10\,$km/s,
but note that other observables may lead to a lower values, closer to $220\pm 20\,$km/s recommended
by the IAU~\cite{Kerr:1986hz}. For the purposes of our estimate,  we assume a central value of $v_c=235$\,km/s, affected by a $\sim 10\%$ error. How reasonable this is can be also inferred from the collection of data shown e.g. in Ref.~\cite{Sofue:2008wt}.

\subsubsection{The halo typical velocity}
The halo velocity distribution cannot be directly observed, but must be inferred from a combination
of observations and theoretical considerations.
In this inference, several sources of error enter. First of all, one assumes that  the {\it rotation curve is constant} over a significant radius around the solar position, i.e. $v_c\simeq const.$ As shown by visual inspection of Fig.~1 in~\cite{Sofue:2008wt}, this approximation holds at not better than $\sim 10\%$ level. For a {\it spherically symmetric system} this implies that the gravitational potential $\Phi$ is logarithmic, since 
\begin{equation}
v_c^2(r)=r\frac{\d \Phi}{\d r} \Rightarrow \Phi (r) =v_c^2 \log(r) +const. 
\end{equation} 
Until now, we assumed nowhere that the DM dominates the overall potential. If we did so, we would find the isothermal profile $\rho\propto r^{-2}$ as unique solution satisfying the above (approximate) observational condition and symmetry condition, and the Maxwell-Boltzmann (MB) distribution $f({\bf v})\propto \exp\left(-\frac{v^2}{v_0^2}\right)$ with $v_0=v_c$ as velocity distribution. However, fit models suggest that DM only contributes a fraction of the $v_c^2$ at the solar system. For examples, the two models in table 2.3 of~\cite{BT} yield  a fraction of 38\% and 64\%. Note also that this fraction is changing with radius, too. So, in order to derive the link between, say, the average dispersion in DM velocity and the DM profile one needs further input, either from fits to the data or simulations.
For a  density of DM which goes as power-law $\rho(r)\propto r^{-\beta}$ in the neighborhood of the Sun, the Eddington equation (see again~\cite{BT}) yields a phase space distribution function for DM which has a dispersion
\begin{equation}
\sigma_v^2\equiv \frac{1}{3}\langle|{\bf v}|^2\rangle=\frac{v_c^2}{\beta}\label{vr:vc}\,.
\end{equation}

it is immediate to check that assuming that the only $v-$dependence enters as $f({\bf v})\propto \exp\left(-\frac{v^2}{v_0^2}\right)$, one gets 
\begin{equation}
\sigma_v^2\equiv\langle v^2\rangle=\int \d^3 {\bf v}\, v^2\,f({\bf v})=\frac{v_0^2}{2}\,\Rightarrow v_0=v_c\sqrt{\frac{2}{\beta}}\,,
\end{equation} 
where the second equality is required for Eq.~(\ref{vr:vc}) to hold. Note that it has been often acritically assumed (perhaps following historical papers as~\cite{Drukier:1986tm} and~\cite{Jungman:1995df}) that $v_0=v_c\,,\label{circ_v0}$ but this strictly requires $\beta= 2$ at the solar radius. If one fits the DM profile with another popular model from numerical simulations, say NFW~\cite{NFW} with a scale radius $\simeq 20\,$kpc, there is a departure from this value of about 20\%. Alone, such uncertainty on the radial profile of DM at the solar radius implies an error of $\sim 10\%$ on the equality $v_c=v_0$. As described previously, a similar level of error appears inherent to the assumption of the constancy of $v_c$.
Other errors, as the limitations of the spherical approximation (and thus isotropy of the space and velocity dependence of the DM profile) are more difficult to assess, but likely present as well, with comparable or larger amplitude. We shall comment on that a bit further in the following. 

In summary, an estimate of the typical velocity parameter $v_0$ is 
\begin{equation}
v_0\approx v_c\,,
\end{equation}
with a $\sim 10\%$ value due to the uncertainty on the measured $v_c$ (and thus correlated with its
true value) and probably an uncertainty {\it at least twice as big} which is inherent to the theoretical
assumptions. The latter errors are unfortunately of systematic nature and hardly reducible in the near future. 

\subsubsection{Escape velocity}

From {\it model-dependent} considerations similar to the ones of the above section one can also derive that expressions for the escape velocity as a function of Galactic parameters, see e.g.~\cite{Drukier:1986tm}. Equivalently, based on milder assumptions and on surveys of high velocity stars,
one can infer constraints such as~\cite{Smith:2006ym}
\begin{equation}
 498\,{\rm km/s}<v_{\rm max}(R_\odot)< 608\,{\rm km/s}\,,
\end{equation}
with a median value of $544\,{\rm km/s}$~\cite{Smith:2006ym} which we adopt in the following as benchmark. Note that since theoretically one expects  that $v_{\rm max}^2\propto v_c^2$, it is not surprising that a similar statistical error of $\sim\,10\%$ is found for both $v_{\rm max}$ and $v_c$. On the other hand,
the local escape velocity depends on {\it global} properties of the dark matter halo. Hence, to be conservative we shall treat the uncertainty on $v_{\rm max}$ as independent from the uncertainty
on $v_0$ as inferred from $v_c$.

\subsubsection{Shape of the velocity distribution}

In the above considerations, for simplicity we parametrized the uncertainty on the velocity distribution via the typical velocity in the halo, or equivalently $v_0$. But the event rates are not only sensitive to the second moment of the distribution, but also to its shape. There is no guarantee that the velocity distribution function is truly MB, and 
actually several numerical simulations (see~\cite{Vogelsberger:2008qb,Zemp:2008gw}) 
suggest that noticeable deviations are present.  To get a feeling for the sensitivity of the
relevant observables to deviations from the MB shape, we parameterize the deviation from a MB. distribution as
\begin{equation}
f({\bf v},\alpha)\propto \left[1+(\alpha-1)\left(\frac{v}{\langle v\rangle}-1\right)\right]\,f_{\rm MB}(v)\label{deviat}
\end{equation}
with $\alpha \in[ 0.7; 1.3]$, as suggested e.g. from the right panel in  Fig. 5 of ~\cite{Vogelsberger:2008qb}.
Note that if the capture equilibration is long, the capture may probe the average distribution,
while the direct detection probes the present one (see below) which is another effect linked to the possible
shape deviation, not explored here.

What we shall employ is the Eq.~(\ref{deviat}), with a MB cut at the $v_{\rm max}$, correctly normalized to one. To that distribution one has then to apply a Galilean transformation to account for the motion of the Sun with respect to the DM halo, see below.

\subsubsection{Motion of the Sun with respect to the DM halo and time-dependent effects}

Another effect on the recoil and capture rate is due to the fact that the Sun is in motion with respect to the DM halo ``rest frame'' with a 
velocity ${\bf v}_\odot$ (since we are interested not in directional signals, only the modulus $v_\odot$ is going to matter in what follows.) Naively, one would expect
\begin{equation}
 v_\odot\approx v_c\,.
\end{equation}

The situation is actually more complicated than that: the Sun is not rotating uniformly around the GC, i.e. it is not at rest with respect to the LSR frame. Then, one has a relative motion between sun and DM halo given by
\begin{equation}
{\bf v}_\odot={\bf v}_{\rm LSR}+{\bf V}_{\rm loc}\,.\label{vodotTOT}
\end{equation}
From the recent analysis in~\cite{Schoenrich:2009bx} one has the relative velocity determination $
{\bf V}_{\rm loc}=\{U,V,W\}=\{11.1\pm0.7\pm1 ,12.2\pm 5\pm 2, 7.3\pm 0.4\pm 0.5\}$km/s respectively
radially inwards ($U$), in the direction of the Galactic rotation ($V$), and vertically upwards ($W$),
with the former error statistical and the second one systematic.
Taken at face value, this only amounts to an upward correction of about $\sim 6\%$ to the naive estimate $v_\odot\approx v_c$, with an uncertainty which is even smaller, thus providing a subleading error.
However, the situation is different when time-dependent effects are taken into account. In fact, as we have seen, one has to contemplate the possibility that the equilibration time for capture in the Sun is comparable to (or longer than) a fraction of its orbital time $\sim 2\times 10^8$ yr. This situation is verified in a significant fraction of SUSY models leading to a large muon flux, as shown in Fig.~\ref{fig:eq} (see also the discussion in Ref. ~\cite{Ellis:2009ka}). For the signal of the Earth, this is true for practically all interesting models.

 \begin{figure*}[th]
\begin{center}
\begin{tabular}{c}
\includegraphics[angle=0,width=0.5\textwidth]{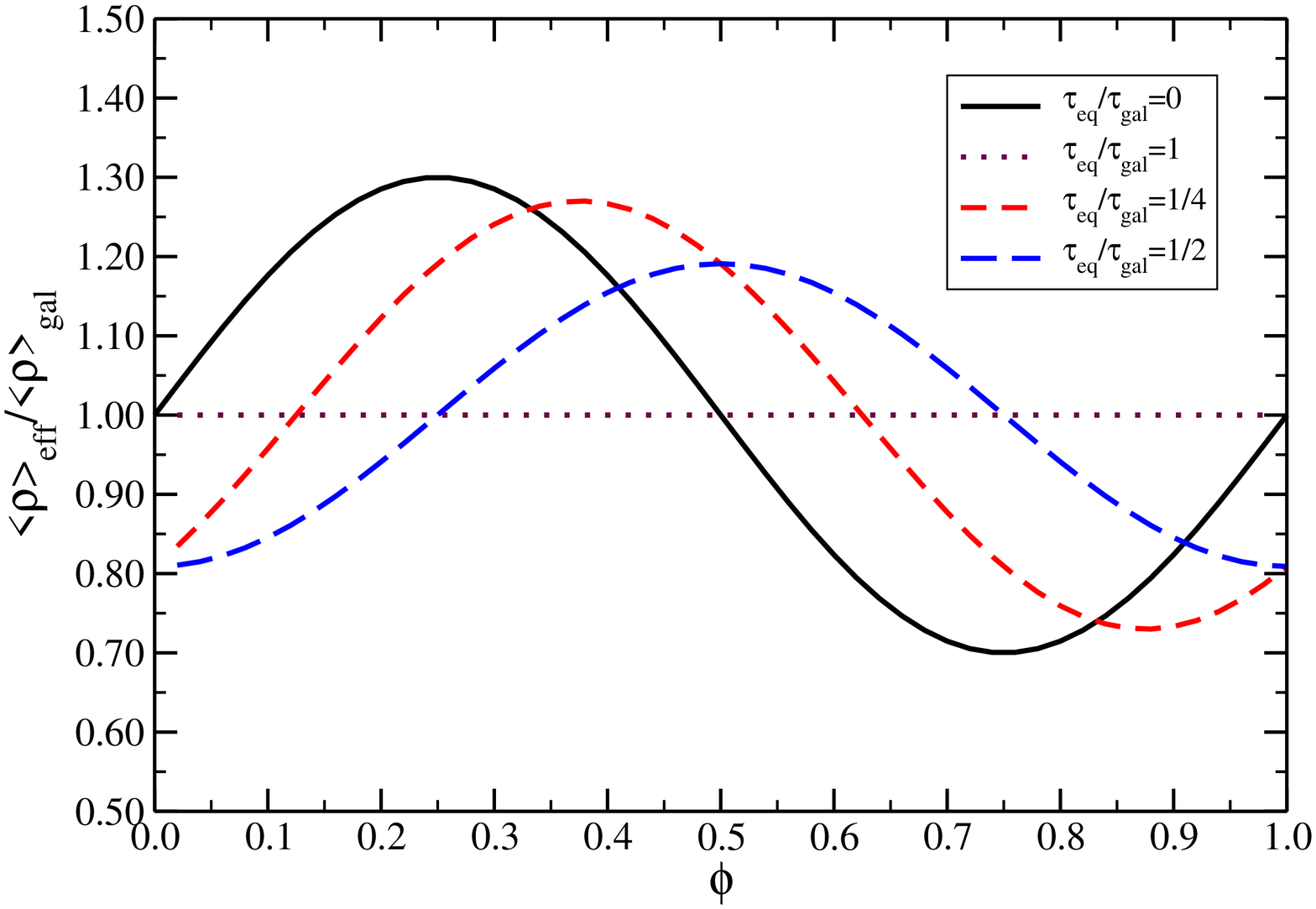}
\includegraphics[angle=0,width=0.5\textwidth]{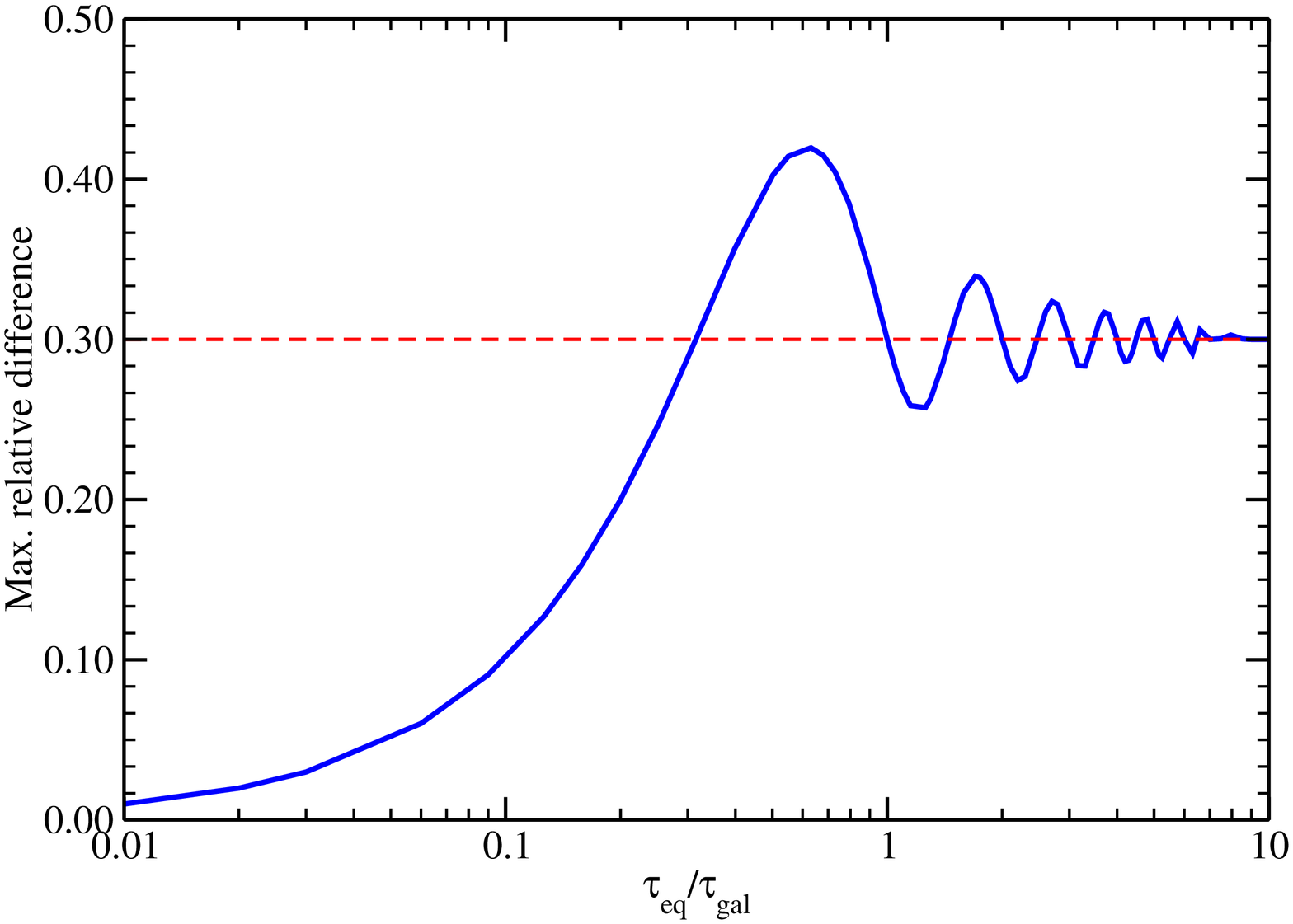}
\end{tabular}
\end{center}
\caption{{\it Left Panel:}  Effective density as a function of the position in a circle of radius $R_\odot$ around the GC, for different choices of $\tau_{eq}$.  {\it Right Panel:} Maximum relative difference between the effective and instantaneous density, as a function of the time to reach equilibrium $\tau_{eq}$, in units of a galactic year $\tau_{gal}$. 
\label{fig:triax}}
\end{figure*}
The fact that the capture signal depends on the long-term past history of the Sun (and Earth) immediately changes the importance of the solar motion effect mentioned  previously: differently from the first term at the RHS of Eq.~(\ref{vodotTOT}), the second term
is time-dependent due to the variation of velocity of the Sun along its orbit in the Galaxy, with
typical timescale of $\sim2\times 10^8$ yr.

To be more quantitative, one would need to know the actual orbit of the Sun in the Galaxy {\it and} the equilibration timescale for the case realized in nature. Both have significant sources of error: i) the error in the ``initial'' (actually final, since the evolution
is performed backwards!) conditions, in particular the velocity vector ${\bf V}_{\rm loc}$; ii) the only approximate knowledge of the Galactic potential; iii) the incomplete knowledge of particle physics and astrophysical parameters determining $\tau_{\rm eq}$. As a result, rather than ``a refinement'' to the naive estimate we can conservatively consider this kind of effects as an additional source of error.
For illustration, we take the results of the numerical integration performed in~\cite{Gies:2005jj}
for the model 2 in~\cite{Dehnen:1997cq} (note that compared with those results the local velocity component has been significantly revised in the analysis reported in~\cite{Schoenrich:2009bx}). As a result, the distance from the GC varies between -2\% and +7\% from
the present value, while the velocity results to anticorrelate almost exactly, with an equal (in modulus) and opposite sign variation (the motion is indeed quasi-keplerian). As a consequence, the two effects almost cancel out for the capture rate in the Sun, so this time-dependence is not the main source of error {\it within the simple model considered}.

The situation is actually different if we move beyond a smooth and isotropic halo approximation:
dropping the isotropy means that the local density might differ from the averaged one (to which
the Solar capture is sensitive for long equilibration times). This difference has been recently evaluated
in~\cite{Zurich10} based on numerical simulations: in the ones including approximatively the effects of baryons, this error is  about $20\%$ to $30\%$. In those without baryons, depending on the ignorance of our orbit within the triaxial halo of DM, the error is as large as a factor $\sim 2$. 
To account for triaxiality, a simple model with a dependence of the density along the Galactic Plane
as  $\rho(\phi)=\langle\rho\rangle_{\rm gal}[1+\varepsilon\sin(2\pi\phi)]$ can be taken as representative of results with baryons, for $\varepsilon=0.3$. Of course, since we do not know the phase at which we are at current time, one can define an effective density
\begin{equation}
\langle\rho\rangle_{\rm eff}(\phi)=\frac{\tau_{\rm gal}}{\tau_{\rm eq}}\int_{-\frac{\tau_{\rm eq}}{\tau_{\rm gal}}}^{0} \rho(\phi+\varphi) \,\d\varphi
\end{equation}
which is probed by the capture rate and is a function of $\phi$. We show in figure~\ref{fig:triax} the effective density as a function of position in a circle of size $R_\odot$ around the GC, for different choices of $\tau_{\rm eq}$ in units of $\tau_{\rm gal}$.  We also show in the same figure (right panel) the maximum relative difference between the effective and instantaneous density, as a function of the time to reach equilibrium $\tau_{\rm eq}$, in units of a galactic year $\tau_{\rm gal}$.  This is obtained at each point by scanning over the unknown $\phi$. In both cases, we fix $\varepsilon=0.3$. Two comments are in order: the maximal error obtained when $\tau_{\rm eq}/\tau_{\rm gal}$ is very large reflects the simple fact that the capture probes the real average value $\langle\rho\rangle_{\rm gal}$, while the recoil the instantaneous value which can differ as much as $\varepsilon=0.3$ from the average (black, solid curve in the left panel of Fig.~\ref{fig:triax}). However, the situation is even more interesting when $\tau_{\rm eq}/\tau_{\rm gal}\alt 0.6$ (which Fig.~1 hints to be a physically  interesting range): in this case the relative error can range from almost negligible to above 40\%, within only a factor 3 variation of $\tau_{\rm eq}$. This suggests that, even if we knew that the ellipticity of the Milky Way halo is at the 30\% level, and even by assuming that an eventual neutrino detection hints at  $\tau_{\rm eq}/\tau_{\rm gal}\alt 1$, the lack of a detailed knowledge of the particle physics and astrophysics limit the estimate of this error conservatively at the 40\% level. Needless to say, should the baryon disk effect be less prominent than what suggested by simulations, this effect could be {\it a factor of several} larger. The actual impact of baryons on the DM structure is in fact still subject to  debate.

\subsubsection{Additional halo (sub)structures}

Concerning the non-smoothness of the halo, an attempt to address its impact on detection rates was performed in Ref.~\cite{Koushiappas:2009ee}, where the effects of the enhanced density while crossing a sub-halo were considered. For our purposes of assessing reasonable errors, rather than exploring what can happen in principle if such an encounter takes place, it is more interesting to assess the probability that such an encounter takes place at all. If we focus on subhalos of mass $M_{\rm sh}=10^{-6} M_\odot$, we can work out the characteristic size, number density therefore average separation for these objects. The Sun encounters a subhalo of mass $M_{\rm sh}$ and size $r_{\rm sh}$ with a frequency $\Gamma_{\rm sh}= 4 \pi v_{\rm tot} n_{\rm sh} r_{\rm sh}^2$. Assuming that a fraction $f$ of the DM is in the form of clumps, and inserting typical values, we find\begin{eqnarray}
\Gamma_{\rm sh} \sim f \,10^{-6} {\rm yr}^{-1}  \frac{v_{\rm sh}}{220 \mbox{km s}^{-1}}  \frac{10^{-6} M_\odot}{M_{\rm sh}}  \left( \frac{r_{\rm sh}}{0.01 \mbox{pc}} \right)^2\label{rateG}
\end{eqnarray}
To perform this estimate $n_{\rm sh}$ we have followed the usual strategy of extrapolating the mass function of subhalos to very small masses, and assumed that the spatial distribution of subhalos traces the smooth DM distribution. The second assumption is most likely a very crude approximation, as the most recent numerical simulation have shown that the distribution of subhalos is {\it anti-biased}, such that if $\rho(r)$ is the total (smooth plus subhalos) DM density profile, $n_{\rm sh}(r)\propto r \rho(r)$.  The fraction $f$ of DM in subhalos is therefore a function of $r$ and it tends to deplete the number of clumps at the solar radius. The results of Ref.~\cite{Diemand:2008in} point towards a more realistic value\footnote{It is perhaps curious to note that if an unrealistically large value of $f$ is used, the actual annihilation signal could be {\it lowered} with respect to its naive equilibrium value in the limit of smooth halo~\cite{Koushiappas:2009ee}.} of $f \sim 10^{-2}$. This estimate suggests that the timescale for an encounter is comparable to (a relevant fraction of) the orbital time of the Sun in the Galaxy. On the other hand, the crossing time $t_{\rm sh}$ for such an encounter is very short, of the order of 
a century. Hence, already on this basis we can conclude that significant alterations of the annihilation
yield, while possible in principle, are very unlikely. In the above considerations, we implicitly assumed that the time during which a significant enhancement of the signal takes place is  comparable with the crossing time. In Appendix~\ref{app1} we show that this is a good assumption if large boosts are required, since a decay time much longer than the crossing time is only possible at the expense of reducing the enhancement in the signal by the same factor. 

We also note that the above estimate is not very dependent on the exact mass of the sub-halo considered: since one has roughly $r_{\rm sh} \sim M_{\rm sh}^{1/3}$, and $t_{\rm sh}\propto r_{\rm sh}$, it follows the weak dependence $\Gamma_{\rm sh} \sim M_{\rm sh}^{-1/3}$, while  $\Gamma_{\rm sh}\,t_{\rm sh}$ which roughly quantifies the probability of ``living during a crossing'' stays virtually constant and very small, of the order of 10$^{-6}$ for the parameters quoted above.

It is worth mentioning that it is not even sufficient that an encounter takes place in order for a significantly enhanced capture to happen. In fact, as a first approximation we can think of each subhalo as orbiting the galaxy with a {\it typical} velocity $w$ sampled from the $f({\bf v})$ distribution. Neglecting the $v_{\rm max}$ cutoff and assuming a MB distribution, the 1D distribution at the Sun is readily written as
\begin{equation}
f_{1}^{{\rm sh}}(w)\approx \frac{2\,w}{\sqrt{\pi}\,v_0\,v_\odot}\sinh\left[\frac{2\,w\,v_\odot}{v_0^2}\right]\,\exp\left(-\frac{w^2+v_\odot^2}{v_0^2}\right)\,.\label{fsh}
\end{equation}
The particles in each sub-halo  have a dispersion $\sigma$ around $w$. However,  as long as $\sigma \ll v_0$ (as expected for the smallest, hence more abundant, sub-halos) this only brings in a small correction. Roughly speaking, as long as long $w< u_{\rm max}$ the DM particles will be captured, when instead in  an encounter one has $w> u_{\rm max}$ only a very small fraction of the DM population could be captured (see Eq.~(\ref{umax})). For fiducial parameters in Eq.~(\ref{fsh}) 
and taking into account that for the Sun $u_{\rm max}\simeq 200\,$km/s for a 100 GeV particle,
one finds that only $\sim 15\%$ of the encounters will lead to a(n efficient) capture. For the Earth, this fraction is negligible.

The scenario could be further enriched if one consider the possible existence of the  ``dark disk''~\cite{Read:2009iv,Bruch:2009rp} due to the dragging effect of the baryonic disk on the DM halo. The crucial aspect is that the (typically subleading by number) population of DM disk particles has a relatively low lag velocity with respect to the stars, $v_{\rm lag}\sim 0-150$ km/s, and a comparably low velocity dispersion $\sigma_d$. As a consequence, the Sun and the Earth effectively see a colder DM gas, a fact that eases captures.  For the Sun, the scaling relation for the capture reported in~\cite{Bruch:2009rp}
suggests that, depending on parameters, the annihilation signal can obtain a correction which ranges
from marginal (a few percent) up to more than one order of magnitude. For the Earth the dependence is
by far more dramatic, usually amounting to orders of magnitude. This confirms once again that the
normalization of the signal from the Earth is extremely dependent on the details of the (very) cold tail of DM phase space distribution and, while still potentially useful for a serendipitous discovery of DM, can tell us very little on the particle physics.

\begin{figure}
\vbox{
\includegraphics[width=1.08\columnwidth]{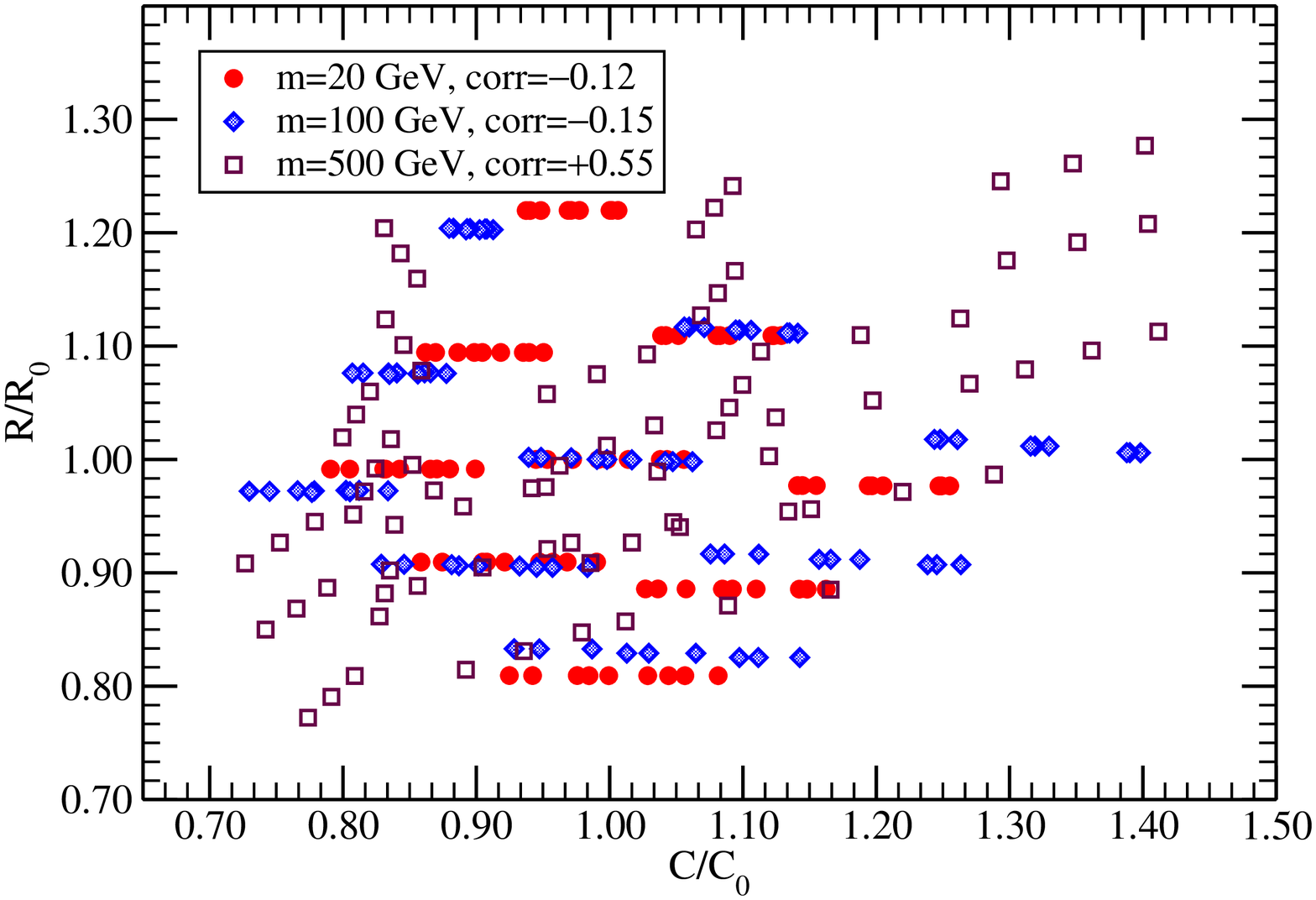}
\vspace*{-0.5cm}
\includegraphics[width=1.08\columnwidth]{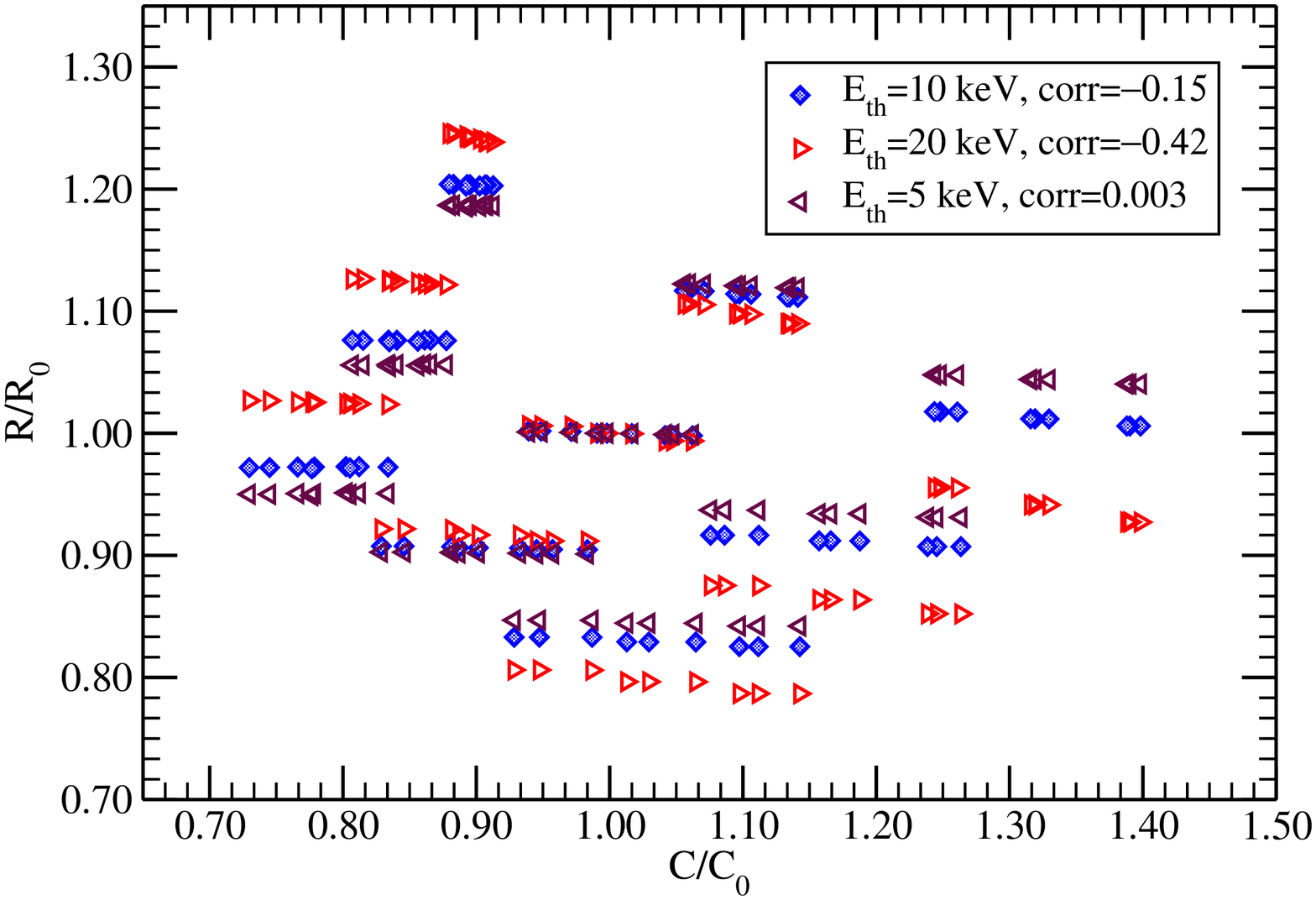}
\vspace*{-0.5cm}
\includegraphics[width=1.08\columnwidth]{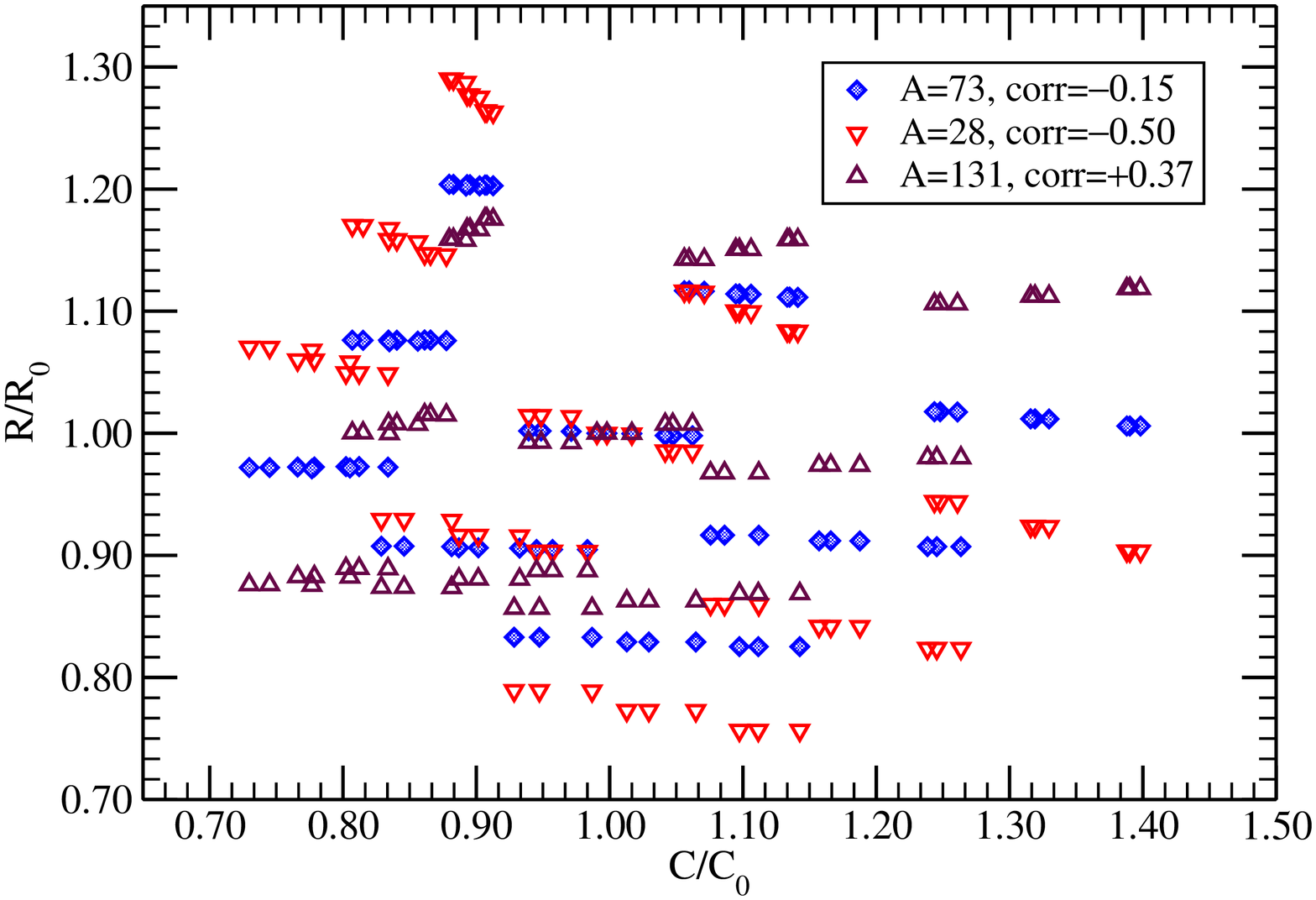}
\caption{Impact of varying simultaneously the four astrophysical parameters ${\bf y}$(see text for range of variation), in the $R-C$ plane, for different values of the DM mass (top) different values of the energy threshold (center) and different values of target mass $A$ (bottom). In each panel, the unmentioned parameters are held fixed at their fiducial values, $m_\chi=100\,$GeV, $E_{\rm th}=10\,$keV, and $A=73$.
\label{fig:dispersion}}
}
\end{figure}

%%%
\section{Halo uncertainties on DM detection rates}
\label{results}
%%%

In most of this section, we estimate the variations in the direct experiment recoil rate $R$ as well as capture rate $C$ when velocity distribution parameters vary within their uncertainties,  as discussed
in Sec. III.1,2,3. At the end of this section, we come back to the additional uncertainties induced by time-dependent effects on the effective density probed, in particular those mentioned in Sec. III.4.

The two functions $R$ and $C$ depend in general from non-astrophysical parameters ${\bf x}$ as well as some astrophysical parameters ${\bf y}$.
Following the discussion in Sec.~\ref{theory}, we select for the former parameters ${\bf x}\equiv\{m_\chi,  E_T, A\} $ denoting respectively: i) the DM mass, which enters both $R$ and $C$, whose fiducial value is put to 100 GeV. ii)  two parameters for the direct detection experiments, namely the threshold energy $E_{\rm th}$ for detectable recoils and the atomic number $A$ of the target material.  Both are model-dependent, but the former is usually in the 10 keV range. We shall assume $E_{\rm th}=10\,$keV
as fiducial value, but we shall explore the consequences for $R$ of varying it by a factor 2,
upwards or downwards.
The target material atomic number $A$  enters our formulae for $C$ via the form-factor dependence. We shall take the fiducial value $A=73$ for an ``intermediate mass'' target, such as Germanium; $A=131$ for a heavy target material as Xenon; $A=28$ for a relatively light target, such as Silicon (these identifications should not be taken too seriously but to investigate the dependence on the target mass, since the form-factor functions are only schematic). 
Note that the DM-direct detector target  cross section (typically spin-independent) $\sigma_{0,A}$ and DM-solar matter target cross section (typically spin-dependent on
hydrogen) $\sigma_{0,p}$ enter just as overall normalizations in $R$ and $C$; thus, exactly like 
the value of the local density of DM $\rho_\odot$, they do not affect our considerations. 

Following the discussion in Sec.~\ref{params}, we identify four main (time-independent)  astrophysical variables ${\bf y}\equiv\{v_0,\,v_{\rm max},\,v_\odot,\,\alpha\}$, denoting respectively the halo dispersion velocity, the escape velocity,  the local velocity of the sun in the DM halo, and the shape-parameter of the velocity distribution.  They are varied as
follows: $v_0$ at fiducial value of 235 km/s as well as varied by $\pm 10\%$, $v_\odot$ at fiducial value of 235 km/s and with $\pm 10\%$ variations, $\alpha$  at the values 0.7, 1, 1.3 and $v_{\rm max}$ is set at 498, 544, and 608 km/s, with the central values of the latter two considered as fiducial ones. 
Note that in particular the assumed 10\% uncertainty on $v_0$ is a very conservative estimate of the error on this quantity, with a more realistic value even a factor of few higher, as discussed in Sec. III.1. This is however consistently with our goal of assessing  a {\it minimal} uncertainty on the observables.

For a first glimpse at the overall dependence of the observables on parameters,
we show in Fig.~\ref{fig:dispersion} a series of ``scatter plots''  in the $R$ vs. $C$ plane scanning over the min-med-max values of the ${\bf y}$ parameters, which are varied simultaneously. These plots are presented for the fiducial values of all but one of the ${\bf x}$-variables: In particular, the  $x$ variable varied in the top panel is the  DM mass, in the central panel it is  the energy threshold for direct detection, and in the bottom panel the target mass $A$ for the direct detection. 
The observables are normalized to the values assumed for fiducial parameters.
It is obvious that, in all cases, the uncertainty on the capture can have excursions of $\sim 40\%$,
while the range span by $R$ is typically within $\pm 20\%$ of its fiducial value, although it can increase for light targets, high thresholds, or heavy particles. What is more important, there are
(anti)correlations  between variations in $R$ and $C$, whose magnitude and even {\it sign} depend
on the variable considered. This is illustrated by diagonal ``trends'' in the dispersion plots,
whose slope can change both quantitatively and in sign. It is not unusual that the relative uncertainty
of the ratio $C/R$, say, is larger than the error on each quantity $C$, $R$ due to this effect. Just for
an illustration of this effect, we computed the correlation coefficient $corr$ of the $R$ 's and $C$'s for the scanned points. This is defined as follows
\begin{equation}
corr(R,C)=\frac{\sum_i (R_i-\langle R\rangle)(C_i-\langle C\rangle)}{\sqrt{\sum_i (R_i-\langle R\rangle)^2\sum_j(C_j-\langle C\rangle)^2}}\,,
\end{equation}
where the index $i$ denotes a particular realization of the arguments, and the sum runs over the
the scanned parameters sets.

 The dateset cross-correlation when varying $m_\chi$=20, 100, 500 GeV varies from $-0.12$ to $-0.15$ and to $+0.55$;
 when varying $E_{\rm th}$=5, 10, 20 keV changes from $0.003$ to $-0.15$
and to $-0.42$;  when varying $A$ from 28 to 73 and 131 rises respectively from $-0.50$ to $-0.15$
and to $+0.37$. Of course, this has no truly statistical meaning (we assumed ``flat'' priors for the variables in their range and did not account for the existing correlations among them), but
it  clearly illustrates the trend of qualitative change of the correlation when underlying parameters
are changed.
 
For a more compact and quantitative exploration of the parameter dependence, it turns useful to define the logarithmic derivatives:
\begin{eqnarray}
\gamma_i ({\bf x},{\bf y}_0)&\equiv& \frac{\partial \ln C({\bf x},{\bf y})}{\partial \ln y_i}\Bigg |_{\bf y=y_0}\,,\\ \varrho_i ({\bf x},{\bf y}_0)&\equiv& \frac{\partial \ln R({\bf x},{\bf y})}{\partial \ln y_i}\Bigg |_{\bf y=y_0}\,.
\end{eqnarray}
The functions $\gamma_i$ and $\varrho_i$ are sufficient (to leading order) in describing the
sensitivities to the quantities of interest. 
Note also that the change in $C/R$ as a consequence of
a change $\delta_i\equiv \delta y_i/y_0$ can be written, in the linear approximation,  as (no sum over
repeated indices)
\begin{equation}
\delta_i \left(\frac{C}{R}\right)\approx\frac{\partial }{\partial y_i}\frac{C}{R}\bigg |_{\bf y_0} \delta y_i=  \frac{C}{R}
\bigg[\gamma_i -\varrho_i\bigg]_{{\bf y}_0}\delta_i\,.
\end{equation}
The values of the $\gamma_i$ and $\varrho_i$ coefficients for fiducial values of ${\bf y}$ and different values of the ${\bf x}$ are reported in Tab.~\ref{tab1} and Tab.~\ref{tab2}.

From these tables it is clear that the dependence on the parameters is actually different in the two cases.  Take the second row of Tab.~\ref{tab1}: it means that an uncertainty in the  velocity of the Sun in the halo is responsible for a factor $|-1.64-0.15|\simeq 1.8$ times larger uncertainty in the {\it relative} normalization of the two rates. 
From that table, one immediately notices the general trend that, when the recoil rate only probes the
high-energy tail of the distribution (light DM particles, high thresholds, light targets) the dependencies
on virtually all parameters are {\it opposite} for $R$ and $C$: the capture in fact always prefers the low-energy tail of the distribution, where particles are slower and thus easier to capture. For higher target masses, a competing effect arises due to the form-factor suppression, so that mixed trends arise. On the other hand, in the limit $E_{\rm th}\to 0$, the dependence on the halo parameters goes to zero since the whole phase space  is actually probed via direct direction.
Note that the effect of the shape parameter $\alpha$ is rather modest, and the relative uncertainties due to the non-MB nature of the distribution can be estimated to be $< 10\%$. A posteriori, this justifies a simple parametric approach to explore this variable.

\begin{table}[t]
\begin{tabular}{|c||c|c||c|c||c|c||}\hline
$i$    & $\gamma_i$ (20) & $\varrho_i$ (20) & $\gamma_i$ (100) & $\varrho_i$ (100)& $\gamma_i$ (500) & $\varrho_i$ (500)\\
\hline\hline 
$v_0$ & $-$0.80& 1.04 &  $ -0.98$  &  $-0.04$    & $ -0.90$  & $-0.22$   \\
\hline
$ v_{\odot}$ & $-$0.94& 1.00 &   $-1.64$  & 0.15&    $-1.92$  & $-0.05$ \\
\hline
$ v_{\rm max}$ &$-$0.12 & 3.21 &   $- 0.13$  &  0.77 &$-0.13$ & 0.50 \\
\hline
$ \alpha$ &$-$0.15 & 0.22 &   $-0.16$  &  0.004 &$-0.12$& $-0.04$ \\
\hline\end{tabular}
\caption{The response functions for capture and direct recoil rate with respect to different
halo parameters for three values of the mass $m_\chi=20,\,100\,,500\,$GeV, and for the fiducial values of the parameters $\alpha=1$, $A=73$, $E_{\rm th}=10\,$keV, $v_c=235\,$km/s, $v_0=235\,$km/s, $v_{\rm max}=544\,$km/s. Note that the product $\sigma\,\rho$ enters only as a scaling, and similarly for both observables (i.e. response is 1).}\label{tab1}
\end{table}

\begin{table}[th]
\begin{tabular}{|c||c|c|c|c||}
\hline
$i$       & $\varrho_i(A=28)$ & $\varrho_i(A=131)$ & $\varrho_i(5\,{\rm keV})$ & $\varrho_i(20\,{\rm keV})$\\
\hline\hline 
$v_0$ &  $ 0.14$  &  $-0.19$   &  $-$0.10  &  0.11   \\
\hline
$ v_{\odot}$ &   $0.32$  & 0.01&    $0.07$  & 0.31\ \\
\hline
$ v_{\rm max}$ &   $1.12$  &  0.52 &$0.69$ & 0.95 \\
\hline
$ \alpha$ &   $0.05$  &  $-0.02$ &$-0.01$& 0.04 \\
\hline
\end{tabular}
\caption{As in Tab.~\ref{tab1}, for the direct detection rate, keeping $m_\chi=100$GeV and varying the
target mass and the threshold.}\label{tab2}
\end{table}

On the top of the different sensitivity to the velocity-distribution variables, as already argued in Sec. III.4, another source of uncertainty is the fact that the two signals sample different {\it time averages} of the physical quantities, in particular the DM density. Accounting for a $\sim$ 40\% maximal error on the ratio of observables estimated on the basis of simulation results (see Fig.~\ref{fig:triax}),  we can conservatively conclude that there is likely a relative uncertainty on $C/R$ which reaches a factor of $\sim 2$, just due to ``halo astrophysics''.

\section{Conclusions}
\label{conclusions}

A conservative estimate of the error on the capture rate suggests that the normalization of the neutrino signal from DM annihilation in the Sun/Earth brings not only the uncertainty given by the local DM density $\rho_\odot$, but often more importantly uncertainties connected with the velocity distribution function, as well as our motion in the halo. We showed that the error on the capture $C$  can easily reach $\approx 40\%$, which translates into a comparable or larger error on the annihilation signal. Even more important, the relative uncertainty on the normalization of the annihilation to recoil signal is typically  larger, easily up to a factor two in presence of  typical equilibration times of $\sim 10^7-10^8$ years\footnote{Going beyond the WIMP paradigm and introducing quite large self-interactions between DM particles can cause a similar mismatch between the two types of observables~\cite{Zentner:2009is}.}. For cases where equilibration times are billions of years or larger (as typical for the Earth) the uncertainties are likely at least few times larger. The same happens if the effects of baryons in reducing Galactic halo triaxiality are smaller than what estimated at present from numerical simulations.

We also studied the sensitivity of the observables to different input parameters, which was shown to be  often opposite for the direct and indirect signals, due to the preferential probe of the high tail part of the velocity distribution for the recoil rate. Interestingly, at least for some variables this anticorrelation can be reversed depending on masses of DM and instrumental parameters.

It is worth noting that these ``halo'' uncertainties appear larger than uncertainties coming from solar composition and/or nuclear/particle physics (for a recent estimate see~\cite{Ellis:2009ka}) and thus provide in most cases the dominant limiting factor in the extracting particle physics information from a future detection of a DM neutrino signal, even when normalizing the rate to some  direct detection event rates. This is true in particular if no or little spectral information is available. Enhancements in the annihilation rate
are a generic expectation of the presence of a ``dark disk'' created by interaction of the halo particles with the baryonic disk, but its effect can range from few percent to orders of magnitude. The role of substructures appears instead marginal at best, barring for highly unlikely circumstances.
In principle, the velocity distribution of WIMP DM can be reconstructed from direct dark matter
  detection data in the range of velocities probed (provided the DM mass is sufficiently constrained), but  $\cal{O}$(100) events are needed to start improving the knowledge of $f_1(v)$ to better than what inferred indirectly via considerations as those reported in Sec.~III (see e.g.~\cite{Drees:2007hr}).

The considerations developed in this article have several implications, whose exploration is left for future works. One obvious consequence is that great care must be paid when comparing the constraints on the spin-dependent scattering cross section inferred from direct detection with those inferred by neutrino experiments (see e.g.~\cite{Flacke:2009eu,Blennow:2009ag,Agrawal:2010ax}), especially when light  DM particles are considered, as well as combinations of experiments with different target masses. This is notably the case of the comparison of the DAMA and CoGent results with the exclusion plots obtained by other experiments (see e.g. Ref. \cite{Savage:2010tg} and references therein).  

Another improvement over the first estimate provided in this article would consist in providing a more statistically sound assessment of the error range in the parameters governing the velocity structure
of the halo (and the motion of the Sun through it), along the lines of what done in~\cite{Catena:2009mf}
for the local halo density. Here we implicitly assumed flat priors in the parameter space and uncorrelated variations of the different parameters. Needless to say, such refinements would become compelling for extracting particle physics parameters if the present generation of WIMP DM detectors were to show any evidence for these particles.

%%%%%%%%%%%%%%%%%%%%%%%%%%%%%%%%%%%%%%%%%%%%%%%%%%%%%%%%%%%%%%%%%%%%%%
\acknowledgments
We would like to thank  F. Iocco and G. Lake  for discussions during the initial stages of this work.

\appendix
\section{Decay time vs. Crossing time}\label{app1}
In Sec. III.5 we have assumed the crossing time of a sub-halo as the relevant timescale for enhancement in the annihilation signal. In principle, however, the relevant quantity is not the time spent by the Sun sitting in a dark-matter subhalo (crossing time), but the time for which the annihilation rate in the Sun is different from the case of a completely smooth halo (enhanced signal decay time). We show here that the latter can only exceed the former at the expense of suppressing the signal enhancement.
 
Let us assume that the Sun/Earth encounters a substructure with density $K\,\rho_\odot$, with $K\gg 1$, during a crossing of time $t_{\rm sh}$ starting at $t=0$. The only case of interest is when a significant capture happens in the passage; in this case, the value reached by $N$ can be significantly larger  than its equilibrium (or long-term) value in the smooth halo, which we can neglect (i.e. $N(0)\approx 0$). Hence, in terms of the capture rate $C$ {\it in the smooth halo}, the approximate solution writes ($\tau_{\rm eq}=(C\,C_A)^{-1/2}$)
\begin{equation}
N(t)\approx\sqrt{K}\sqrt{\frac{C}{C_A}}\tanh\left(\frac{t\sqrt{K}}{\tau_{\rm eq}}\right)\,.
\end{equation}
As a consequence, the enhancement in the annihilation signal at the end of the crossing over the ``naive'', long-term average signal in the smooth halo is given by the following ``boost'' function
\begin{equation}
B(t_{\rm sh})=\frac{\Gamma_A(t_{\rm sh})}{\Gamma_0(t_\odot)}\approx K\frac{\tanh^2\left(\frac{t_{\rm sh }\sqrt{K}}{\tau_{\rm eq}}\right)}{\tanh^2\left(\frac{t_{\odot}}{\tau_{\rm eq}}\right)}\,,
\end{equation}
where the equilibration time is the one defined with respect to the smooth halo.

For $t> t_{\rm sh}$, the enhanced capture rate ends, and the enhanced signal starts to 
decline. Its time evolution is now dictated by
\begin{equation}
\dot N\approx-C_A\,N^2\,,
\end{equation}
hence, for $t>t_{\rm sh}$,
\begin{equation}
N(t_{\rm sh})-N(t)=N(t_{\rm sh})N(t)\,C_A(t-t_{\rm sh})\,.
\end{equation}

The signal (proportional to $N^2$) will drop to a factor $1/F$ of the one at $t_{\rm sh}$ after a time delay $\Delta_F t$ given by
\begin{equation}
\Delta_F t=\frac{\sqrt{F}-1}{C_AN(t_{\rm sh})}=\frac{(\sqrt{F}-1)\tau_{\rm eq}}{
\sqrt{K}\tanh\left(\frac{t_{\rm sh}\sqrt{K}}{\tau_{\rm eq}}\right)}\,.
\end{equation}
Ideally, to maximize both the boost and the duration of the enhanced signal one would require that both $B(t_{\rm sh})$ and $\Delta_F t$ are as large as possible (note that for any practical circumstance $(\sqrt{F}-1)\sim {\mathcal O}(1)$). Clearly, these two requests are in tension with each other. There are two limiting situations (for simplicity, we assume in the following that the smooth halo signal has reached equilibrium): 

For equilibrium to be reached during the crossing, which guarantees the ``full boost'', we require
\begin{equation}
t_{\rm sh }\gtrsim \tau_{\rm eq}/\sqrt{K}\,,
\end{equation}
hence
\begin{equation}
B(t_{\rm sh})\approx K\,,\:\:\:\:\:\:\Delta_F t\approx\frac{(\sqrt{F}-1)\tau_{\rm eq}}{
\sqrt{K}}\lesssim t_{\rm sh} \,.
\end{equation}
So, the optimal case for enhanced signal requires a ``decay time'' after crossing which is comparable or more rapid than the crossing time.

If instead equilibrium is not reached, assuming the argument of the `$\tanh$' function to be small one obtains
\begin{eqnarray}
B(t_{\rm sh})&=&K\epsilon\,, \:\:\:\epsilon\approx K\left(\frac{t_{\rm sh }}{\tau_{\rm eq}}\right)^2\ll 1\,,\\
\Delta_F t &\approx &\frac{(\sqrt{F}-1)}{\epsilon}\,t_{\rm sh}\,.
\end{eqnarray}

More in general, one has
\begin{equation}
B(t_{\rm sh})\frac{\Delta_F t}{t_{\rm sh}}\approx(\sqrt{F}-1) K\frac{\tanh(y)}{y}\lesssim (\sqrt{F}-1) K\,
\end{equation}
where $y=\sqrt{K}\,t_{\rm sh}/\tau_{\rm eq}$. The inequality is saturated only for small $y$, while the expression at the LHS is suprressed as $1/y$ for large $y$.
So, we conclude that a decay time much longer than the crossing time is only possible at the expense of reducing the enhancement in the signal by the same factor.

%%%%

%%%%%%%%%%%%%%%%%%%%%%%%%%%%%%%%%%%%%%%%%%%%%%%%%%%%%%%%%%%%%%%%%%%%%%

\begin{thebibliography}{00}

\bibitem{Jungman:1995df}
  G.~Jungman, M.~Kamionkowski and K.~Griest, 
  %`Supersymmetric dark matter', 
  {\it Phys.\ Rept.\ } {\bf 267} (1996) 195.

%
\bibitem{Bergstrom:2000pn}
  L.~Bergstrom,
  %``Non-baryonic dark matter: Observational evidence and detection methods,''
  Rept.\ Prog.\ Phys.\  {\bf 63} (2000) 793.
%  [arXiv:hep-ph/0002126].
  %%CITATION = RPPHA,63,793;%%

\bibitem{Bertone:2004pz}
  G.~Bertone, D.~Hooper and J.~Silk,
  %``Particle dark matter: Evidence, candidates and constraints,''
  Phys.\ Rept.\  {\bf 405} (2005) 279.
%   [arXiv:hep-ph/0404175].

\bibitem{book}
{\it Particle Dark Matter: Observations, Models and Searches}, ed. G. Bertone, 2010, Cambridge University Press

%\cite{Baltz:2006fm}
\bibitem{Baltz:2006fm}
  E.~A.~Baltz, M.~Battaglia, M.~E.~Peskin and T.~Wizansky,
  %``Determination of dark matter properties at high-energy colliders,''
  Phys.\ Rev.\  D {\bf 74} (2006) 103521
  [arXiv:hep-ph/0602187].
  %%CITATION = PHRVA,D74,103521;%%

\bibitem{Nath:2010zj}
  P.~Nath {\it et al.},
  %``The Hunt for New Physics at the Large Hadron Collider,''
  Nucl.\ Phys.\ Proc.\ Suppl.\  {\bf 200-202} (2010) 185.
 % [arXiv:1001.2693 [hep-ph]].
  %%CITATION = NUPHZ,200-202,185;%%
%
\bibitem{Bertone:2010rv}
  G.~Bertone, D.~G.~Cerdeno, M.~Fornasa, R.~R.~de Austri and R.~Trotta,
  %``Identification of Dark Matter particles with LHC and direct detection
  %data,''
  arXiv:1005.4280 [hep-ph].
  %%CITATION = ARXIV:1005.4280;%%

  \bibitem{Bertone:2007xj}
  G.~Bertone, D.~G.~Cerde\~no, J.~I.~Collar and B.~C.~Odom,
  %``WIMP identification through a combined measurement of axial and scalar
  %couplings,''
  Phys.\ Rev.\ Lett.\  {\bf 99} (2007) 151301.
  %[arXiv:0705.2502 [astro-ph]].
%\cite{}

\bibitem{Drees:2008bv}
  M.~Drees and C.~L.~Shan,
  %``Model-Independent Determination of the WIMP Mass from Direct Dark Matter
  %Detection Data,''
  JCAP {\bf 0806} (2008) 012.
%  [arXiv:0803.4477 [hep-ph]].
  %%CITATION = JCAPA,0806,012;%%
  %\cite{Peter:2009ak}
  
\bibitem{Peter:2009ak}
  A.~H.~G.~Peter,
  %``Getting the astrophysics and particle physics of dark matter out of
  %next-generation direct detection experiments,''
  Phys.\ Rev.\  D {\bf 81} (2010) 087301
  [arXiv:0910.4765 [astro-ph.CO]].
  %%CITATION = PHRVA,D81,087301;%%

\bibitem{Shan:2010qv}
  C.~L.~Shan,
  %``Effects of Residue Background Events in Direct Dark Matter Detection
  %Experiments on the Reconstruction of the Velocity Distribution Function of
  %Halo WIMPs,''
  JCAP {\bf 1006} (2010) 029
  [arXiv:1003.5283 [astro-ph.HE]].
  %%CITATION = JCAPA,1006,029;%%
  


  
%
\bibitem{Profumo:2010ya}
  S.~Profumo and P.~Ullio,
  %``Multi-wavelength Searches for Particle Dark Matter,''
  arXiv:1001.4086 [astro-ph.HE].
  %%CITATION = ARXIV:1001.4086;%%


  %\cite{Gould:1992}
\bibitem{Gould:1992}
  A.~Gould,
  %``Cosmological density of WIMPs from solar and terrestrial annihilations,''
   Astrophys.\ J.\  {\bf 388}, 338 (1992).

  
  %\cite{Kuijken:1989hu}
  
  
\bibitem{dyn1}
  J.~Holmberg and C.~Flynn,
  %``The local surface density of disc matter mapped by Hipparcos,''
  Mon.\ Not.\ Roy.\ Astron.\ Soc.\  {\bf 352} (2004) 440
  [arXiv:astro-ph/0405155].
  %%CITATION = MNRAA,352,440;%%;
%\cite{Holmberg:1998xu}
\bibitem{dyn2}
  J.~Holmberg and C.~Flynn,
  %``The local density of matter mapped by Hipparcos,''
  Mon.\ Not.\ Roy.\ Astron.\ Soc.\  {\bf 313} (2000) 209
  [arXiv:astro-ph/9812404].
  %%CITATION = MNRAA,313,209;%%  
  \bibitem{dyn3}
K.~Kuijken and G.~Gilmore, ApJ {\bf 267} (1991)  L9 
\bibitem{dyn4}
  K.~Kuijken and G.~Gilmore,
  %``The Mass Distribution in the Galactic Disc - Part Two - Determination of
  %the Surface Mass Density of the Galactic Disc Near the Sun,''
  Mon.\ Not.\ Roy.\ Astron.\ Soc.\  {\bf 239} (1989) 605.
  %%CITATION = MNRAA,239,605;%%

\bibitem{Catena:2009mf}
  R.~Catena and P.~Ullio,
  %``A novel determination of the local dark matter density,''
  arXiv:0907.0018 [astro-ph.CO].

\bibitem{Strigari:2009zb}
  L.~E.~Strigari and R.~Trotta,
  %``Reconstructing WIMP Properties in Direct Detection Experiments Including
  %Galactic Dark Matter Distribution Uncertainties,''
  JCAP {\bf 0911} (2009) 019.
%  [arXiv:0906.5361 [astro-ph.HE]].
  %%CITATION = JCAPA,0911,019;%%
  
  \bibitem{Salucci:2010qr}
  P.~Salucci, F.~Nesti, G.~Gentile and C.~F.~Martins,
  %``The dark matter density at the Sun's location,''
  arXiv:1003.3101 [astro-ph.GA].
  %%CITATION = ARXIV:1003.3101;%%

  \bibitem{Zurich10} 
  %\cite{Pato:2010yq}
  M.~Pato, O.~Agertz, G.~Bertone, B.~Moore and R.~Teyssier,
  %``Systematic uncertainties in the determination of the local dark matter
  %density,''
  arXiv:1006.1322 [astro-ph.HE].
  %%CITATION = ARXIV:1006.1322;%%

\bibitem{Ellis:2009ka}
  J.~Ellis, K.~A.~Olive, C.~Savage and V.~C.~Spanos,
  %``Neutrino Fluxes from CMSSM LSP Annihilations in the Sun,''
  Phys.\ Rev.\  D {\bf 81}, 085004 (2010).
 % [arXiv:0912.3137 [hep-ph]].
  %%CITATION = PHRVA,D81,085004;%%

\bibitem{Belli:2002yt}
  P.~Belli, R.~Cerulli, N.~Fornengo and S.~Scopel,
  %``Effect of the galactic halo modeling on the DAMA/NaI annual modulation
  %result: an extended analysis of the data for WIMPs with a purely
  %spin-independent coupling,''
  Phys.\ Rev.\  D {\bf 66}, 043503 (2002).
%  [arXiv:hep-ph/0203242].
  %%CITATION = PHRVA,D66,043503;%%

\bibitem{Kuhlen:2009vh}
  M.~Kuhlen {\it et al.},
  %``Dark Matter Direct Detection with Non-Maxwellian Velocity Structure,''
  JCAP {\bf 1002}, 030 (2010).
%  [arXiv:0912.2358 [astro-ph.GA]].
  %%CITATION = JCAPA,1002,030;%%

  \bibitem{Vogelsberger:2008qb}
  M.~Vogelsberger {\it et al.},
  %``Phase-space structure in the local dark matter distribution and its
  %signature in direct detection experiments,''
  arXiv:0812.0362 [astro-ph].
  %%CITATION = ARXIV:0812.0362;%%
  
\bibitem{Belanger:2008sj}
  G.~Belanger, F.~Boudjema, A.~Pukhov and A.~Semenov,
  %``Dark matter direct detection rate in a generic model with micrOMEGAs2.1,''
  Comput.\ Phys.\ Commun.\  {\bf 180}, 747 (2009).
%  [arXiv:0803.2360 [hep-ph]].
  %%CITATION = CPHCB,180,747;%%


\bibitem{Alenazi:2006wu}
  M.~S.~Alenazi and P.~Gondolo,
  %``Phase-space distribution of unbound dark matter near the Sun,''
  Phys.\ Rev.\  D {\bf 74}, 083518 (2006).
  %[arXiv:astro-ph/0608390].
  %%CITATION = PHRVA,D74,083518;%%

\bibitem{Lundberg:2004dn}
  J.~Lundberg and J.~Edsjo,
  %``WIMP diffusion in the solar system including solar depletion and its
  %effect on earth capture rates,''
  Phys.\ Rev.\  D {\bf 69}, 123505 (2004).
%  [arXiv:astro-ph/0401113].
  %%CITATION = PHRVA,D69,123505;%%

%\cite{Peter:2009mi}
\bibitem{Peter:2009mi}
  A.~H.~G.~Peter,
  %``Dark matter in the solar system I: The distribution function of WIMPs at
  %the Earth from solar capture,''
  Phys.\ Rev.\  D {\bf 79}, 103531 (2009).
%  [arXiv:0902.1344 [astro-ph.HE]].
  %%CITATION = PHRVA,D79,103531;%%


%\cite{Peter:2008sy}
\bibitem{Peter:2008sy}
  A.~H.~G.~Peter and S.~Tremaine,
  %``Dynamics of WIMPs in the solar system and implications for detection,''
  PoS {\bf IDM2008}, 061 (2008).
%  [arXiv:0806.2133 [astro-ph]].
  %%CITATION = POSCI,IDM2008,061;%%
%\bibitem{Peter:2009mk}
  A.~H.~G.~Peter,
  %``Dark matter in the solar system II: WIMP annihilation rates in the Sun,''
  Phys.\ Rev.\  D {\bf 79}, 103532 (2009).
%  [arXiv:0902.1347 [astro-ph.HE]].
  %%CITATION = PHRVA,D79,103532;%%
  
\bibitem{serenelli}
 \texttt{http://www.mpa-garching.mpg.de/$\sim$aldos/solar\_main.html}

%\cite{Dogan:2010my}
\bibitem{Dogan:2010my}
  G.~Dogan, A.~Bonanno and J.~Christensen-Dalsgaard,
  %``Near-surface effects and solar-age determination,''
  arXiv:1004.2215 [astro-ph.SR].
  %%CITATION = ARXIV:1004.2215;%%

  \bibitem{earth}D. Alfe, M. J. Gillan, and G. D. Price (2003)
%   A Thermodynamics from first principles: temperature and composition of the Earth's core.
Mineralogical Magazine {\bf 67}, 113-123.

\cite{Mack:2007xj}
\bibitem{Mack:2007xj}
  G.~D.~Mack, J.~F.~Beacom and G.~Bertone,
  %``Towards Closing the Window on Strongly Interacting Dark Matter:
  %Far-Reaching Constraints from Earth's Heat Flow,''
  Phys.\ Rev.\  D {\bf 76} (2007) 043523.
  [arXiv:0705.4298 [astro-ph]].
  %%CITATION = PHRVA,D76,043523;%%
  
  
\bibitem{Gould:1987ir}
  A.~Gould,
  %``Resonant Enhancements In Wimp Capture By The Earth,''
  Astrophys.\ J.\  {\bf 321}, 571 (1987).
  %%CITATION = ASJOA,321,571;%%
  
   
  %\cite{Gondolo:2004sc}
\bibitem{Gondolo:2004sc}
  P.~Gondolo, J.~Edsjo, P.~Ullio, L.~Bergstrom, M.~Schelke and E.~A.~Baltz,
  %``DarkSUSY: Computing supersymmetric dark matter properties numerically,''
  JCAP {\bf 0407} (2004) 008.
%  [arXiv:astro-ph/0406204].
  %%CITATION = JCAPA,0407,008;%%
  
  %\cite{Halzen:2009vu}
\bibitem{Halzen:2009vu}
  F.~Halzen and D.~Hooper,
  %``The Indirect Search for Dark Matter with IceCube,''
  New J.\ Phys.\  {\bf 11}, 105019 (2009).
 % [arXiv:0910.4513 [astro-ph.HE]].
  %%CITATION = NJOPF,11,105019;%%

\bibitem{BT}
  J.~Binney and S.~Tremaine,
 ``Galactic Dynamics'', Princeton Series in Astrophysics, 2nd edition (2008).
   
 
	
   
\bibitem{Kerr:1986hz}
  F.~J.~Kerr and D.~Lynden-Bell,
  %``Review of galactic constants,''
  Mon.\ Not.\ Roy.\ Astron.\ Soc.\  {\bf 221}, 1023 (1986).
  %%CITATION = MNRAA,221,1023;%%

\bibitem{Sofue:2008wt}
  Y.~Sofue, M.~Honma and T.~Omodaka,
  %``Unified Rotation Curve of the Galaxy -- Decomposition into de Vaucouleurs
  %Bulge, Disk, Dark Halo, and the 9-kpc Rotation Dip --,''
  arXiv:0811.0859 [astro-ph].
  %%CITATION = ARXIV:0811.0859;%
  
     \bibitem{Drukier:1986tm}
  A.~K.~Drukier, K.~Freese and D.~N.~Spergel,
  %``Detecting Cold Dark Matter Candidates,''
  Phys.\ Rev.\  D {\bf 33}, 3495 (1986).
  %%CITATION = PHRVA,D33,3495;%%
      
      
    \bibitem{NFW} J. F. Navarro {\it et al.},
Astrophys. J. 462 563 (1996).


  
  \bibitem{Smith:2006ym}
  M.~C.~Smith {\it et al.},
  %``The RAVE Survey: Constraining the Local Galactic Escape Speed,''
  Mon.\ Not.\ Roy.\ Astron.\ Soc.\  {\bf 379}, 755 (2007).
%  [arXiv:astro-ph/0611671].
  %%CITATION = MNRAA,379,755;%%
  


%\cite{Zemp:2008gw}
\bibitem{Zemp:2008gw}
  M.~Zemp {\it et al.},
  %``The Graininess of Dark Matter Haloes,''
  arXiv:0812.2033 [astro-ph].
  %%CITATION = ARXIV:0812.2033;%%

\bibitem{Schoenrich:2009bx}
  R.~Schoenrich, J.~Binney and W.~Dehnen,
  %``Local Kinematics and the Local Standard of Rest,''
  arXiv:0912.3693 [astro-ph.GA].
  %%CITATION = ARXIV:0912.3693;%%
  
\bibitem{Gies:2005jj}
  D.~R.~Gies and J.~W.~Helsel,
  %``Ice Age Epochs and the Sun's Path Through the Galaxy,''
  Astrophys.\ J.\  {\bf 626}, 844 (2005).
 % [arXiv:astro-ph/0503306].
  %%CITATION = ASJOA,626,844;%%

\bibitem{Dehnen:1997cq}
  W.~Dehnen and J.~Binney,
  %``Local stellar kinematics from Hipparcos data,''
  Mon.\ Not.\ Roy.\ Astron.\ Soc.\  {\bf 298}, 387 (1998).
 % [arXiv:astro-ph/9710077].
  %%CITATION = MNRAA,298,387;%%
  
\bibitem{Koushiappas:2009ee}
  S.~M.~Koushiappas and M.~Kamionkowski,
  %``Galactic Substructure and Energetic Neutrinos from the Sun and the Earth,''
  Phys.\ Rev.\ Lett.\  {\bf 103}, 121301 (2009).
%  [arXiv:0907.4778 [astro-ph.CO]].
  %%CITATION = PRLTA,103,121301;%%
  
  
%\cite{Diemand:2008in}
\bibitem{Diemand:2008in}
  J.~Diemand, M.~Kuhlen, P.~Madau, M.~Zemp, B.~Moore, D.~Potter and J.~Stadel,
  %``Clumps and streams in the local dark matter distribution,''
  Nature {\bf 454} (2008) 735.
%  [arXiv:0805.1244 [astro-ph]].
  %%CITATION = NATUA,454,735;%%


%\cite{Read:2009iv}
\bibitem{Read:2009iv}
  J.~I.~Read, L.~Mayer, A.~M.~Brooks, F.~Governato and G.~Lake,
  %``A dark matter disc in three cosmological simulations of Milky Way mass
  %galaxies,''
  arXiv:0902.0009 [astro-ph.GA].
  %%CITATION = ARXIV:0902.0009;%%


\bibitem{Bruch:2009rp}
  T.~Bruch, A.~H.~G.~Peter, J.~Read, L.~Baudis and G.~Lake,
  %``Dark Matter Disc Enhanced Neutrino Fluxes from the Sun and Earth,''
  Phys.\ Lett.\  B {\bf 674}, 250 (2009).
%  [arXiv:0902.4001 [astro-ph.HE]].
  %%CITATION = PHLTA,B674,250;%%
  
\bibitem{Zentner:2009is}
  A.~R.~Zentner,
  %``High-Energy Neutrinos From Dark Matter Particle Self-Capture Within the
  %Sun,''
  Phys.\ Rev.\  D {\bf 80}, 063501 (2009).
 % [arXiv:0907.3448 [astro-ph.HE]].
  %%CITATION = PHRVA,D80,063501;%%

\bibitem{Drees:2007hr}
  M.~Drees and C.~L.~Shan,
  %``Reconstructing the velocity distribution of WIMPs from direct dark matter
  %detection data,''
  JCAP {\bf 0706}, 011 (2007).
%  [arXiv:astro-ph/0703651].
  %%CITATION = JCAPA,0706,011;%%


\bibitem{Flacke:2009eu}
  T.~Flacke, A.~Menon, D.~Hooper and K.~Freese,
  %``Kaluza-Klein Dark Matter And Neutrinos From Annihilation In The Sun,''
  arXiv:0908.0899 [hep-ph].
  %%CITATION = ARXIV:0908.0899;%%

\bibitem{Blennow:2009ag}
  M.~Blennow, H.~Melbeus and T.~Ohlsson,
  %``Neutrinos from Kaluza-Klein dark matter in the Sun,''
  JCAP {\bf 1001}, 018 (2010).
%  [arXiv:0910.1588 [hep-ph]].
  %%CITATION = JCAPA,1001,018;%%


\bibitem{Agrawal:2010ax}
  P.~Agrawal, Z.~Chacko, C.~Kilic and R.~K.~Mishra,
  %``Direct Detection Constraints on Dark Matter Event Rates in Neutrino
  %Telescopes, and Collider Implications,''
  arXiv:1003.5905 [hep-ph].
  %%CITATION = ARXIV:1003.5905;%%

%\cite{Savage:2010tg}
\bibitem{Savage:2010tg}
  C.~Savage, G.~Gelmini, P.~Gondolo and K.~Freese,
  %``XENON10/100 dark matter constraints in comparison with CoGeNT and DAMA:
  %examining the Leff dependence,''
  arXiv:1006.0972 [astro-ph.CO].
  %%CITATION = ARXIV:1006.0972;%%



\end{thebibliography}
\end{document}